%% file: main.tex
\documentclass[11pt,letterpaper]{article}

\usepackage{graphicx}
\usepackage{amssymb}
\usepackage{amsthm}
\usepackage{amsmath}
\usepackage[margin=1in]{geometry}
\usepackage{placeins}
\usepackage[font={small,it}]{caption}
\usepackage{subfigure}
\usepackage{authblk}
\usepackage{algorithm,algorithmic}
\usepackage{hyperref}

\usepackage{titletoc}
\title{California Reservoir Drought Sensitivity and Exhaustion Risk using Statistical Graphical Models}

\newcommand{\data}{\mathcal{D}}
\newcommand{\train}{\mathrm{train}}
\newcommand{\test}{\mathrm{test}}
\newcommand{\Sp}{\mathbb{S}}
\newcommand{\suchthat}{\mathrm{s.t.}}
\newcommand{\R}{\mathbb{R}}
\newcommand{\given}{\mid}

\newcommand{\argmin}{\operatornamewithlimits{arg\,min}}
\newcommand{\tr}{\operatorname{tr}}
\newcommand{\E}{\operatorname{\mathbb{E}}}
\newcommand{\eqnarrpad}{\hspace{-4ex}} 
\date{December 2, 2015}

\author[a,1]{Armeen Taeb}
\author[b]{J. T. Reager} 
\author[b]{Michael Turmon}
\author[a]{Venkat Chandrasekaran}

\affil[a]{California Institute of Technology, Pasadena, CA, 91125}
\affil[b]{Jet Propulsion Laboratory, California Institute of Technology, Pasadena, CA, 91109}

\begin{document}

%
%

\maketitle

\begin{abstract}
The ongoing California drought has highlighted the potential vulnerability of state water management infrastructure to multi-year dry intervals. Due to the high complexity of the network, dynamic storage changes across the California reservoir system have been difficult to model using either conventional statistical or physical approaches. Here, we analyze the interactions of monthly volumes in a network of 55 large California reservoirs, over a period of 136 months from 2004 to 2015, and we develop a latent-variable graphical model of their joint fluctuations. We achieve reliable and tractable modeling of the system because the model structure allows unique recovery of the best-in-class model via convex optimization with control of the number of free parameters. We extract a statewide `latent' influencing factor which turns out to be highly correlated with both the Palmer Drought Severity Index (PDSI, $\rho \approx 0.86$) and hydroelectric production ($\rho \approx 0.71$). Further, the model allows us to determine system health measures such as exhaustion probability and per-reservoir drought sensitivity. We find that as PDSI approaches -6, there is a probability greater than 50\% of simultaneous exhaustion of multiple large reservoirs.\end{abstract}

\input{intro} 
\input{data}  
\input{method} 
\input{results} 
\input{discuss} 





\section{Acknowledgement} This research was partly carried out at the Jet Propulsion Laboratory, California Institute of Technology, under contract with NASA.


\clearpage
\newpage
\newpage
\input{Appendix}

\end{document}

%% file: intro.tex
%
%


\section{Introduction}
\newcommand\dropcap{\empty}

\dropcap
The state of California depends on a complex water management system to meet wide-ranging water demands across a large, hydrologically diverse domain.  As part of this infrastructure, California has constructed 1530 reservoirs, and a collective storage capacity equivalent to a year of mean runoff from California rivers \cite{graf1999dam}.  The purpose of this system is to create water storage capacity and extend seasonal water availability to meet national agricultural, residential, industrial, power generation, and recreational needs.

Major statewide California precipitation deficits during water years 2012–2015 rivaled the most intense 4-year droughts in the past 1200 years \cite{griffin2014unusual}. The drought was punctuated by low snowpack in the Sierra Nevada, declining groundwater storage, fallowed agricultural lands, and lower reservoir storage \cite{aghakouchak2014global, famiglietti2014global,howitt2014economic}.  How long these conditions will persist is unknown, and recent research has explored whether their origin is ascribable to natural climate variability alone or has been increasingly exacerbated by anthropogenic influence  \cite{diffenbaugh2015anthropogenic,williams2015contribution,shukla2015temperature}.  
A strong 2015–2016 El Nino was expected to bring some relief for drought conditions \cite{hoell2015does}, but the 2016 precipitation through Spring was not enough to substantially decrease statewide drought severity.  These events illustrate that there exists some sensitivity of California reservoirs to drought conditions, with resulting implications for national water and agricultural security.  Yet this sensitivity has been difficult to quantify due to the size and complexity of the reservoir network.

Recent research has explored models to determine the potential for future exhaustion of US reservoirs such as Lake Mead (e.g., \cite{barnett:dry-2008,barnett2009sustainable}).  Such approaches yield potentially valuable information to water managers, but the underlying approach used to statistically
model a single reservoir is much simpler than an approach for modeling large network of hundreds of reservoirs, whose complex management is based on multiple economic and sectoral objectives \cite{howitt2014economic}. 
The hard-to-quantify influence of human operators and lack of system closure
have made the modeling and prediction of reservoir storage behavior using physical equations challenging, for example in hydrology or climate models \cite{solander2016simulating}. 
Thus, it is difficult to say with any certainty how the complex network of California reservoirs responds hydrologically to drought severity, 
and at what point exhaustion becomes feasible.

We present a statistical approach to the problem of modeling California’s complex reservoir network. We model the network variables with a statistical graphical model and use the historical system behavior to estimate network interactions by solving a constrained maximum-likelihood estimation problem.  
The robustness and predictive accuracy of the model is greatly improved by imposing sparsity constraints; the form of these constraints allows application of new convex optimization tools to uniquely solve the resulting estimation problem. 

In its general form, 
the model allows automatic recovery of 
unobserved latent variables that influence 
network behavior, as well as introduction of key forcing variables,
most significantly, drought severity index.
This approach allows for a mathematical simplification 
of a complex network so that network behavior and response to forcing 
can be better understood than before.
The learned model allows us to make statistically significant statements about the sensitivity of a representative subset of 
California’s reservoirs to drought severity, 
and the resulting conditions leading 
to increased probability of exhaustion in the future.

%% file: data.tex
%
%


\section{Data}
\label{sec:data}

Our primary dataset is monthly averages of reservoir
volumes,
derived from daily time series of volumes 
downloaded from the California Data Exchange Center (CDEC).
We also used secondary data for some covariates.

\subsection{Reservoir Time Series}


We began with a list of all 60 California reservoirs
having daily data during the period of study
(January 2004 -- April 2015).
From this list, we excluded the five reservoirs with 
more than half of their values undefined or zero,
leaving 55 reservoirs.
This list of daily values was inspected using a simple continuity 
criterion and 
approximately 50 
specific values were removed or corrected.
Corrections were possible in six cases because values 
had misplaced decimal points, but all other detected
errors were
set to missing values.
The most common error modes were missing values that were recorded as 
zero volume, and a burst of errors in the Lyons reservoir
during late October 2014 that seem due to a change in
recording method at that time.

The final set of 55 reservoir volume 
time series spans 4595 days over the
136 months in the study period.
It contains two full cycles of 
California drought (roughly, 2004--2010 and 2010--present).
Four California hydrological zones are represented, with 
25, 20, 6, and 4
reservoirs in the Sacramento, San Joaquin, Tulare,
and North Coast zones, respectively.


The volumetric data was averaged from daily
down to monthly values,
and its strong annual component 
was removed with a per-month, per-station average, leaving
a volume anomaly with $136$ monthly observations to be modeled statistically.
Before being used in the fitting algorithms of
Section~\ref{sec:method}, each time series is also
rescaled by its standard deviation so that each series has 
unit variance. 
The volume anomaly, being a de-trended average of small
daily fluctuations, is plausibly modeled
by a Gaussian frequency law.

\subsection{Covariate Time Series}

Our model and analysis makes use of ancillary data, i.e., \emph{covariates}, which are observable variables, exogenous to the model,
that may typically affect a large fraction of reservoirs.
The particular covariates we used are temperature, 
snowpack, PDSI, number of agricultural workers, hydroelectric power, and consumer price index (CPI).
We defined metrics for
statewide snowpack (in high-altitude regions) as well as zone-specific snowpack. We apply a time lag of two months to the covariates temperatures, snowpack, and PDSI. As with the reservoir time series, 
we remove seasonal patterns with a per-month average, 
and scale each time series to have unit variance. 


%% file: method.tex
%
%



\section{Methods}
\label{sec:method}

We introduce three modeling techniques that are employed on the reservoir dataset.
We denote the $55$ reservoir volumes by 
$y \in \R^{55}$, and the nine covariates described in Section~\ref{sec:data} by 
$c \in \R^{9}$. We denote the $136$ monthly observations by 
$\data_n = \{y^{(i)}\}_{i = 1}^n \subset \R^{55}$. In what follows, we assume zero-mean variables,
or equivalently, that
any known mean has been subtracted from the observations ---
assured in our case by preprocessing that subtracts a climatology
from $y$ and $c$.
\subsection{Graphical modeling} 
\begin{figure}
\centering
\includegraphics[width=4.5in]{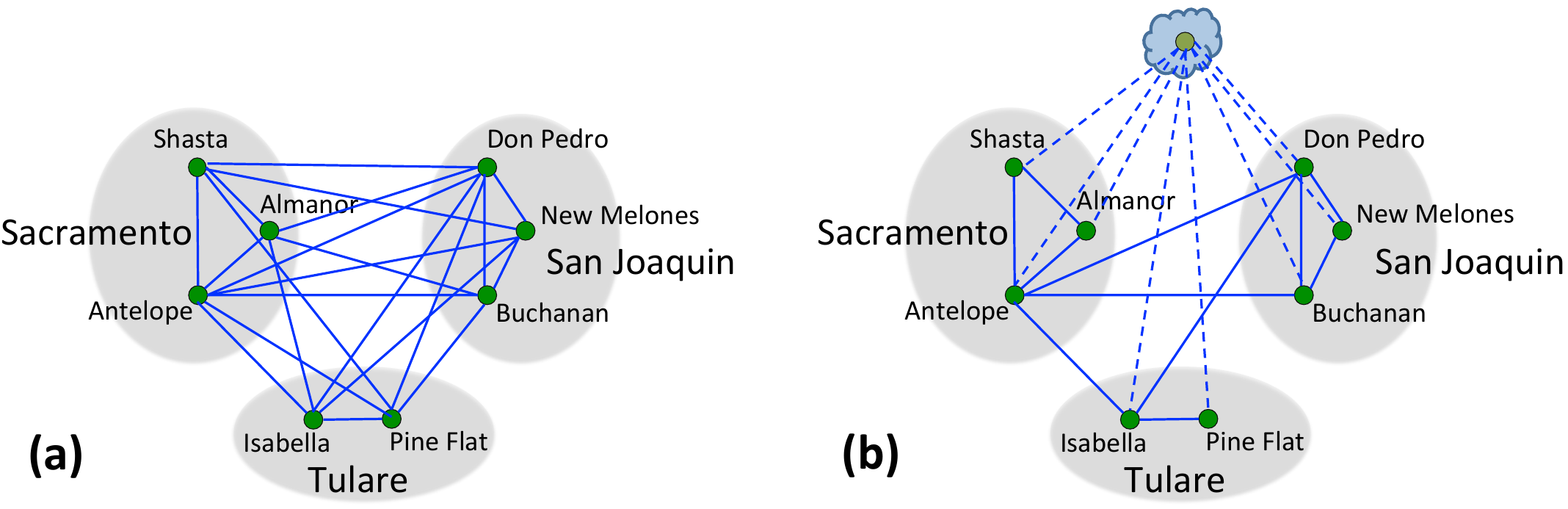}%
\caption{Graphical structure between a collection of reservoirs -- without latent variables (a) -- with latent variables (b). Green nodes represent reservoirs (variables) and the clouded green node represents latent variables. Solid blue lines represent edges between reservoirs and dotted edges between reservoirs and latent variables. The reservoirs have been grouped according to hydrological zones.}
\label{fig:graph-model-ex}
\end{figure}

A natural framework for modeling a large network of reservoirs is a Gaussian graphical model (defined with respect to a graph) where nodes index reservoirs and edges denote dependencies between reservoirs. The absence of an edge between a pair of reservoirs implies that the corresponding reservoir volumes are independent conditioned on the volumes of the remaining reservoirs in the network. Figure~\ref{fig:graph-model-ex}(a) is an example of a graphical model over a collection of reservoirs in three hydrological zones: Sacramento, San Joaquin, and Tulare. In comparison to graphical modeling, standard linear regression techniques suffer from over-fitting in this setting where there is a moderate number of observations compared to the number of reservoirs. We demonstrate this point in the supporting information (SI).

The structural properties of a graphical model are encoded 
in the sparsity pattern of the inverse covariance matrix 
(the \emph{precision matrix}) over reservoirs. 
Specifically, consider a Gaussian graphical model for reservoir volumes $y$ with a 
covariance matrix $\Sigma$.
(Both $\Sigma$ and the corresponding precision matrix $\Theta = \Sigma^{-1}$ 
must be symmetric, written $\Theta \in \Sp^{55}$,
and positive definite, written $\Theta \succ 0$.)
Then, an edge between reservoirs $r$ and $r'$ is present in the graph
if and only if $\Theta_{r,r'} \neq 0$,
with larger magnitudes indicating stronger interactions. Equivalently, the volumes of reservoirs $r$ and $r'$ are conditionally independent if and only if $\Theta_{r,r'} = 0$. Thus, learning a Gaussian graphical model
is equivalent to estimating a \emph{sparse} precision matrix $\Theta$~\cite{Yuan,Friedman}. 


\subsection{Latent variable graphical modeling} 
A drawback with graphical modeling is that we do not observe all relevant phenomena: some globally-influential quantities may be
\emph{hidden} or \emph{latent} variables. Without accounting for these latent variables, there will be confounding dependencies between many reservoirs. For example, water held by a 
collection of nearby reservoirs might be influenced by a common snowpack variable.
Without observing this common variable, all reservoirs in this set would 
appear to have mutual links, whereas if snowpack is included in the analysis, the
common behavior is explained by a link to the snowpack variable.
Accounting for latent structure removes these confounding dependencies and leads to sparser interactions between reservoirs. Figure~\ref{fig:graph-model-ex}(b) illustrates this point.
Recent work \cite{Chand2012} developed a principled approach, 
termed latent-variable graphical modeling,
to account for latent variables.
Fitting a latent-variable graphical model corresponds to 
representing $\Theta$ (the precision matrix of the reservoir volumes) 
as the sum of a sparse and a low-rank matrix, 
rather than just a sparse matrix, 
as in the case of pure graphical modelling. 
Here the low-rank component of $\Theta$ accounts for the effect of the latent variables, and its rank is the number of latent variables. The sparse component of $\Theta$ specifies the conditional graphical model structure underlying the distribution of reservoir volumes conditioned on the latent variables. 

Based on this decomposition of $\Theta$, a natural model selection method for learning a latent-variable graphical model is via a regularized maximum-likelihood estimator. Specifically, we fit the $n$ observations $\data_n = \{y^{(i)}\}_{i = 1}^n \subset \R^{55}$ to a latent-variable graphical model using the estimator:
\begin{eqnarray}
    (\hat{S}, \hat{L}) = \argmin_{S,\,L \in \Sp^{55}} 
    &&
    -\ell(S-L; \data_n) +
    \lambda (\gamma \|S\|_{1}+ \tr(L)) \nonumber 
    \\ 
    \suchthat 
    &&
    S - L \succ 0,\, L \succeq 0 
\quad.
\label{eqn:LVmodeling}
\end{eqnarray}
The term $\ell(\Theta; \data_n)$ is the Gaussian log-likelihood 
of $\Theta$, which, after discarding additive constants and scaling, is
the concave function
\begin{equation}
    \ell(\Theta; \data_n) 
   = 
    \log\det(\Theta) - 
    \tr \left[\Theta \cdot \tfrac{1}{n}\sum_{i=1}^n y^{(i)}{y^{(i)}}' \right]
    \quad .
\label{eqn:loglike}
\end{equation}
The constraints $ \succ 0$ and $ \succeq 0$ 
impose positive-definiteness and positive semi-definiteness.  
Here $\hat{S}$ provides an estimate for the sparse component of $\Theta$ (corresponding to the conditional precision matrix of reservoir volumes), and $\hat{L}$ provides an estimate for the low-rank component of $\Theta$ (corresponding to the effect of latent variables on reservoir volumes). 
We note that constraining $L$ to be the zero matrix in \eqref{eqn:LVmodeling} 
provides an estimator for learning a sparse Gaussian graphical model; see \cite{Friedman} for further details.
The function $\|\cdot\|_{1}$ denotes the
$L_1$ norm (element-wise sum of absolute values) that promotes sparsity in the matrix $S$. 
This regularizer has successfully been employed in many settings for fitting structured sparse models to high-dimensional data (see the survey \cite{Wainwright} and the references therein). 
The role of the trace penalty on $L$ is to promote low-rank structure \cite{Fazel}.
The {regularization parameter} $\gamma$ provides a trade-off between the graphical model component and the latent component, and the regularization
parameter $\lambda$ provides overall control of the
trade-off between the fidelity of the model to the data 
and the complexity of the model. 
For $\lambda, \gamma \geq 0$, the regularized maximum likelihood estimator 
of \eqref{eqn:LVmodeling} is a convex program with a unique optimum.  
Further theoretical support for this estimator is presented in \cite{Chand2012}.

\subsection{Latent Variable Structure} 
Latent variable graphical modeling identifies the effect that underlying latent variables have on the reservoir volumes. For applications, the following approach to associate semantics to these latent variables can aid understanding of the estimated model. 
Suppose we learn a latent-variable graphical model with $k$ latent variables with the methodology just described. 
Let $z \in \R^k$ denote the latent variables and $y \in \R^{55}$ denote reservoir volumes; further, partition the joint precision matrix of $(y,z)$ as $\tilde{\Theta} = \begin{pmatrix} \tilde{\Theta}_y & \tilde{\Theta}_{zy}' \\ \tilde{\Theta}_{zy} & \tilde{\Theta}_z \end{pmatrix}$. 
A natural approximation for the time series of $z$ given observations $\mathcal{D}_n$ is 
the conditional mean:
\begin{equation}
   \tilde{z}^{(i)} 
  =
  \E [ z^{(i)} \given y^{(i)} ] 
  = 
  \tilde{\Theta}_{z}^{-1}\tilde{\Theta}_{zy}y^{(i)}.
\label{eqn:ts}
\end{equation}
If $\tilde{\Theta}_{z}$ and $\tilde{\Theta}_{zy}$ were explicitly known, the 
length-$n$ time series
$\{\tilde{z}^{(i)}\}_{i=1}^n \subset \R^{k}$ would provide an estimate of the latent variables
given observations $\mathcal{D}_n$. 
As discussed in \cite{Chand2012}, the low-rank component in the decomposition of the marginal precision matrix of $y$ is 
$L = \tilde{\Theta}_{zy}' \tilde{\Theta}_{z}^{-1}\tilde{\Theta}_{zy}$. 
However, having an estimate for $L$ does not uniquely identify
$\tilde{\Theta}_{z}^{-1}\tilde{\Theta}_{zy}$.
Indeed, for any non-singular $A \in \R^{k \times k}$, one can transform 
$\tilde{\Theta}_z \to A\tilde{\Theta}_z{A}'$ and $\tilde{\Theta}_{zy} \to A\tilde{\Theta}_{zy}$ without altering $L$. 
In terms of $z$,
these observations imply that for any non-singular $A$, 
$\{{A}^{-1}\tilde{z}^{(i)}\}_{i=1}^n$ is 
an equivalent realization of the latent variable time series:
$z$ is recoverable only up to a nonsingular transformation.

Nevertheless, the structure of the low-rank matrix $L$ places a constraint on the effect of the latent variables $z$ on $y$.
Let $\tilde{Z} \in \R^{n \times k}$ denote a (non-unique) 
realization of latent variable observations. 
As we have seen, $\tilde{Z}{A'}^{-1}$ is an equivalent realization. 
The key \emph{invariant} is the column-space of $\tilde{Z}$,
a $k$-dimensional linear subspace of $R^{n}$, 
which we term the \emph{latent space}, which can be recovered as follows.

Letting $Y \in \R^{n\times 55}$ denote observations of reservoir volumes, \eqref{eqn:ts} becomes 
$\tilde{Z} = Y\tilde{\Theta}_{zy}' \tilde{\Theta}_{z}^{-1}$. 
Since the column-space of $Y\tilde{\Theta}_{zy}' \tilde{\Theta}_{z}^{-1}$ is equal to the column-space of $YL$, one
set of basis elements for the latent space is the 
$k$ left singular vectors of the matrix $Y{L}$, which can be readily computed.
We interpret the underlying latent variables by linking observed covariates to this latent space.
Specifically, we evaluate the correlation of a covariate with the latent variables by computing the energy of the projection of the time series of the covariate onto the latent space. Covariates with large correlation represent interpretable components of the underlying latent variables:
these covariates have significant overlap with the latent space.

\subsection{Conditional latent-variable graphical modeling}
\label{sec:lvgm-covariate}
We provided an approach for linking a
latent variable space defined by $L$ to observed covariates. We can extend our modeling framework to  incorporate covariates $x \in \R^q$ that are most correlated with the latent space, where $q$ is the number of covariates included from the overall collection $c \in \mathbb{R}^9$ described in Section 1. Since the covariates $x$ can account for the effect of some of the latent variables 
in the previous modeling framework, the distribution of $y$ given $x$ may still depend on
a few \emph{residual latent variables}. Therefore, we fit a latent-variable graphical model to the conditional distribution of $y\given x$.

In the SI, we discuss how the convex program 
of \eqref{eqn:LVmodeling} 
can be modified for fitting a conditional latent-variable graphical model to the augmented data
$\data_n^{+} = \{ (y^{(i)},\, x^{(i)})\} \subset \R^{55+q}$. 
This estimator produces three components: 
$\hat{\Theta}$, $\hat{L}_y$, and $\hat{S}_y$.
The matrix $\hat{\Theta} \in \Sp^{55+q}$ 
is an estimate for 
the joint precision matrix of $(y,x)$, 
and in particular, its submatrix 
$\hat{\Theta}_y$ is an estimate for the conditional precision matrix 
of the reservoirs given the covariates. The submatrix $\hat{\Theta}_y$ can be decomposed as $\hat{\Theta}_y = \hat{S}_y - \hat{L}_y$. The matrix
$\hat{L}_y \in \Sp^{55}$ is an estimate for the effect of residual latent variables on 
$y \given x$, with the rank of $\hat{L}_y$ equal to number of residual latent variables.
Finally, $\hat{S}_y \in \Sp^{55}$ is an estimate of 
the precision matrix of $y \given x$ conditioned on these residual latent variables.

%% file: results.tex
%
%


\section{Results}
\label{sec:results}

We now apply the methods described above to the reservoir system. 
For our experiments, we set aside monthly observations 
of reservoir volumes and covariates from 
August 2005 -- August 2013 as a training set ($n_{\train} = 96$) and monthly observations 
from January 2004 -- July 2005 and September 2013 -- April 2015 
as a (disjoint) testing set ($n_{\test} = 40$).  
Let $\mathcal{D}_{\train}$ and $\mathcal{D}_{\train}^{+}$, respectively,
be the training set of reservoir volumes and that of
reservoir volumes augmented with covariate data; 
$\mathcal{D}_{\test}$ and $\mathcal{D}_{\test}^{+}$ are 
the corresponding testing sets.  The particular covariates that are included in the augmented datasets $\mathcal{D}^+_{\mathrm{train}}$ and $\mathcal{D}^+_{\mathrm{test}}$ are discussed later.

Recall that, for each variable, 
we remove seasonal components by subtracting a per-month average
as computed on the training set, leaving an anomaly that we model as Gaussian.
We scale time series to have unit variance on the training observations,
and then apply the same transformation to the testing set. The convex programs associated with each modeling framework are regularized max-det programs and are solved in polynomial time using general-purpose solvers \cite{Toh}. 


\subsection{Graphical Modeling} 
\begin{figure}[thbp]
\centering
\bfseries{\sffamily{Reservoir Connectivity}}
\includegraphics[width = 5in, height = 4.5in]{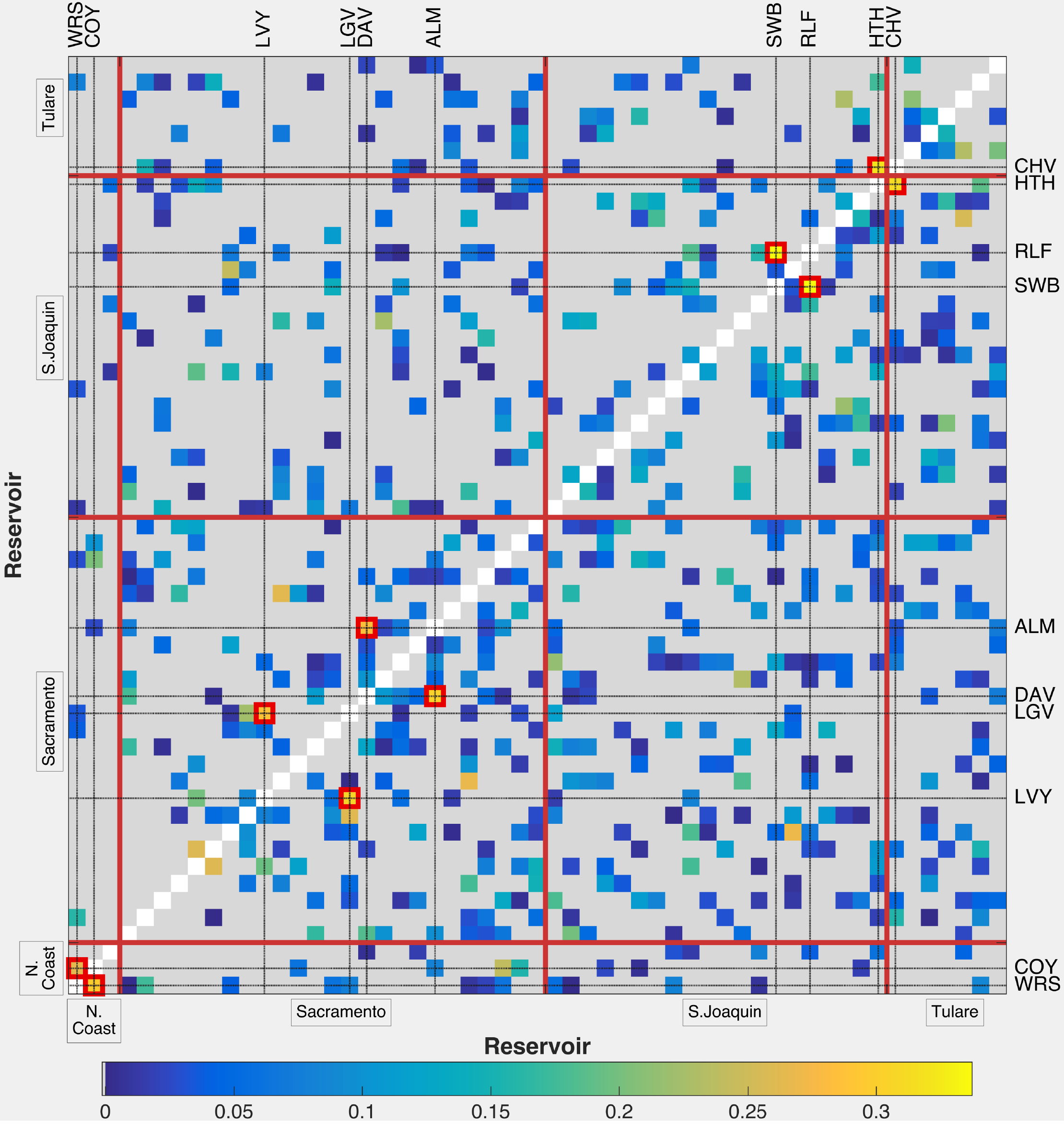}
\caption{Linkages between reservoir pairs
in the latent-variable sparse graphical model (upper triangle, 2 latent variables) compared with those of the ordinary sparse model (lower triangle). 
Connection strength $s(r,r')$ is shown in the image map,
with unlinked reservoir pairs drawn in gray. 
The four hydrological zones are separated by red lines.
Red boxes surround the five strongest connections in each model.}
\label{fig:graphicalmodel2}
\end{figure}

We begin by learning a graphical model over the reservoirs in our network. 
Specifically, we use $\mathcal{D}_{\train}$ as input to the appropriate convex program --- this is the program \eqref{eqn:LVmodeling} with $L$ constrained to be the zero matrix.
We select the single regularization parameter $\lambda$ ($\gamma$ is set to $1$ without loss of generality) by cross-validation with $\mathcal{D}_{\mathrm{test}}$ 
to learn a graphical model consisting of $385$ edges. 
Recall from Section~\ref{sec:method} that the sparsity pattern of the precision matrix $\hat{\Theta}$ encodes the dependence structure between reservoirs,
so that a zero entry in the precision matrix implies 
that the corresponding reservoir volumes are independent conditioned on the volumes of the remaining reservoirs in the network. We denote the strength of an edge as the 
normalized magnitude of the corresponding precision matrix entry, that is,
\begin{equation*}
  s(r,r') 
  = 
  {|\Theta_{r,r'}|}\big/(\Theta_{r,r}\Theta_{r',r'})^{1/2}
  \ge 0 
  \quad.
\end{equation*}
The lower triangle of Figure~\ref{fig:graphicalmodel2} shows the dependence relationships between reservoirs in this graphical model. 
The five strongest edges in this graphical structure are between reservoirs 
Relief--Main Strawberry, Cherry--Hetch Hetchy, Little Grass Valley--Lake Valley, Almanor--Davis, and Coyote Valley--Warm Spring. 
The presence of these strong edges is sensible as each such edge is between reservoirs in the same hydrological zone, 
and $4$ of these $5$ edges are between pairs of reservoirs fed by the same river basin.

We also observe that the majority of interactions in this graphical model are among reservoirs that have similar volumetric capacities. Specifically, $80\%$ of the edges are between reservoirs that are within a factor of $5$ in capacity. As a point of comparison, consider a complete graph with all pairwise interactions. In this graphical structure, $55\%$ of the edges are among reservoirs that are within a factor of $5$ in capacity. Thus, graphical modelling removes the dependencies between reservoirs of vastly different capacity. This is expected since reservoirs with substantially different volumetric capacity are unlikely to have similar variability. In a similar vein, we notice that a significant portion of the edges in the graphical model are between reservoirs at similar elevations. For example, of the ten strongest edges, nine are between reservoirs that are within a factor of $1.7$ in altitude. 

Finally, we observe that a large portion of the strong interactions occur between reservoirs in the same hydrological zone, here denoted $g(r)$.
To quantify this observation, we consider 
\begin{equation*}
    \kappa = 
    \frac{\sum_{r \ne r' \mathrm{~and~} g(r) = g(r')} s(r,r') }
    {\sum_{r \ne r'} s(r,r')}
\quad,
\end{equation*}
the ratio of within-zone edge strength to total edge strength.
The model we fit has $\kappa = 0.72$, so 72\% of interactions
were within-zone.
Nevertheless, we notice that there are a number of strong edges between reservoirs that are geographically far apart.
This suggests the presence of global state-wide latent variables 
(Figure~\ref{fig:graph-model-ex}(b)) that influence distant reservoirs, and accounting for these variables may remove confounding relationships between geographically distant reservoirs.

\subsection{Latent-Variable Graphical Modeling}

We use the method described in Section~\ref{sec:method} 
to account for the effect of latent variables when fitting a graphical model. 
Specifically, using observations 
$\mathcal{D}_{\train}$ 
as input to the convex program \eqref{eqn:LVmodeling},
and an appropriate choice of regularization parameters $(\lambda, \gamma)$ 
(chosen via cross-validation with $\mathcal{D}_{\test}$), 
we learn a latent-variable graphical model with estimates $\hat{S}$ and $\hat{L}$. Recall that $\hat{L}$ represents the effect of the latent variables on reservoir volumes and its rank is the number of latent variables; 
$\hat{S}$ is the conditional precision matrix of reservoir volumes conditioned on the latent variables, 
with its sparsity pattern encoding the conditional graphical structure among reservoirs. The best model consists of two latent variables, together with a conditional graphical model (conditioned on the latent variable) having $259$ edges.

The upper triangle portion of Figure~\ref{fig:graphicalmodel2} shows the dependency relationships between reservoir pairs in the sparse component of the latent-variable graphical model. 
Comparing this graphical structure with the graphical structure without latent
variables, we observe that incorporating latent variables 
weakens or removes many connections between reservoirs:
$124$ are removed and $232$ are weakened by including the latent variables. 
Of the $124$ edges removed, $79$ are between reservoirs in different hydrological zones. Further, the ratio of inner zone edge strengths to total edge strength increases from 
$\kappa = 72\%$ to $\kappa = 79\%$. These results support the idea that latent variables extract global features that are common to all reservoirs, and incorporating them results in more localized interactions.

Finally, of the $55$ reservoirs in our system, $35$ are used for sourcing hydroelectric power. In the graphical structure without latent variables, there are $155$ pairwise edges between reservoirs that are used for generating hydroelectric power. Once the latent variables are incorporated, all but $8$ of these edges are weakened or removed from the graphical structure. This suggests that hydroelectric power may be strongly correlated to the latent variables. We verify this hypothesis next.

\subsection{Latent-Variable Structure}

\begin{figure}[t]
\centering
\bfseries{\sffamily{Palmer Drought Index Comparison}}
\includegraphics[width=4.5in,height=2.4in]{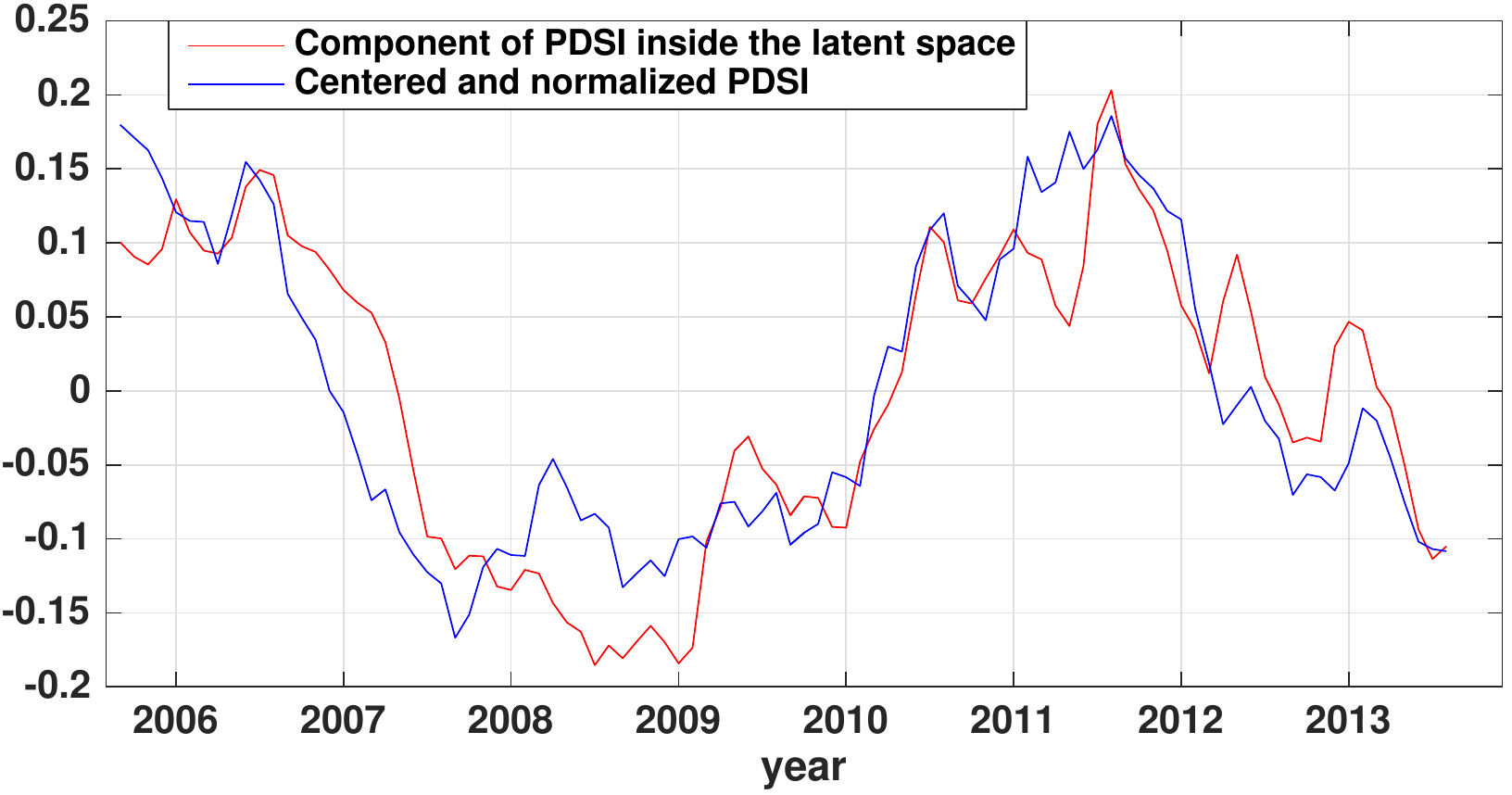}
\caption{Time series of the component of PDSI inside the latent space compared with PDSI itself. 
Both time series are centered and scaled to have unit variance.}
\label{fig:time}
\end{figure}

We associate semantics to the latent variables learned in the latent-variable graphical model. Following the approach presented in Section~\ref{sec:method}, we correlate each of the $9$ covariates described in Section $1$ to the two-dimensional latent space.
\footnote{Recall from Section~\ref{sec:data} that we apply a time lag of two months to the covariates PDSI, snowpack (all zones), and temperature (all basins).} 
The covariates for Palmer drought severity index (PDSI) and hydroelectric power are dominant with correlation values of $\rho = 0.86$ and $\rho = 0.71$. Secondary covariate influences are due to snowpack, with little influence from temperature. 
The complete list of correlations 
is in Table~\ref{table:covariate-relevance}.
Figure~\ref{fig:time} gives 
a side-by-side comparison of PDSI and the component of PDSI inside the latent space. 


\begin{table}[h]
\centering 
\begin{tabular}{lc}
\hline\hline 
Covariate & Correlation \\[0.1ex]
\hline 
Palmer drought severity index (PDSI) & $0.86$\\
Hydroelectric power & $0.71$ \\
Snow pack: San Joaquin zone & $0.46$ \\
Snow pack: Tulare zone  & $0.38$\\
Snow pack: Sacramento zone & $0.28$ \\
Consumer price index (CPI) & $0.24$ \\
Number of agricultural workers & $0.22$ \\
Temperature: San Joaquin basin & $0.14$ \\
Temperature: Sacramento basin & $0.07$ \\
\hline 
\end{tabular}
\vspace{0.2in}
\caption{Covariates and correlations with the latent space} 
\label{table:covariate-relevance} 
\end{table}


\vspace{-0.35in}
\subsection{Sensitivity to Drought and Exhaustion Probability}
Having discovered PDSI 
as a variable that influences the reservoir system state-wide, we use this covariate to characterize the system-wide response to drought. 
Our measure for the behavior of a reservoir is 
\emph{exhaustion probability}:
the probability that a reservoir volume drops to zero. To obtain a system-wide response to drought, we compute the probability of exhaustion of a collection of reservoirs conditioned on particular PDSI. We obtain this probability by learning a joint distribution over reservoir volumes and PDSI. We compute this probability for the month of August,
when reservoirs are typically at their lowest, but the same calculation applies to any month. Since we applied a time lag of two months to the PDSI time series, these  probabilities are computed based on June PDSI.

To learn a joint distribution, 
let $x \in \mathbb{R}$ denote PDSI and consider a 
conditional latent-variable graphical model over $(y,x) \in \mathbb{R}^{55+1}$. 
Using observations $\mathcal{D}_{\text{train}}^{+}$ (consisting of $55$ reservoir volumes and PDSI values) as input to the convex program 
described in section~\ref{sec:lvgm-covariate},
we choose parameters $(\lambda, \gamma)$ via cross-validation with $\mathcal{D}^+_{\test}$
to fit a latent-variable graphical model to the conditional distribution $y~|~x$. 
The estimated model consists of $1$ residual latent variable.  Recall that fitting a latent-variable graphical model to the reservoir volumes $y$ (without conditioning on the PDSI covariate) resulted in a model with $2$ latent variables.  Evidently, by regressing away the effect of PDSI, we are left with one residual latent variable, which supports the observation that PDSI is highly correlated with the latent space in the latent-variable graphical model of $y$ (without PDSI).  The conditional latent-variable graphical modeling procedure also provides an estimate of a graphical model of the conditional distribution of $y$ conditioned on $x$ (PDSI) as well as the one residual latent variable -- this graphical model consists of $274$ edges.  Finally, we obtain an estimate $\hat{\Theta} \in \mathbb{S}^{56}$ of the joint precision matrix over $(y,x)$.

\begin{figure}[h]
\centering
\bfseries{\textsf{~~ ~~System-Wide Response to Drought}}
\includegraphics[width=4.5in,height=2.2in]{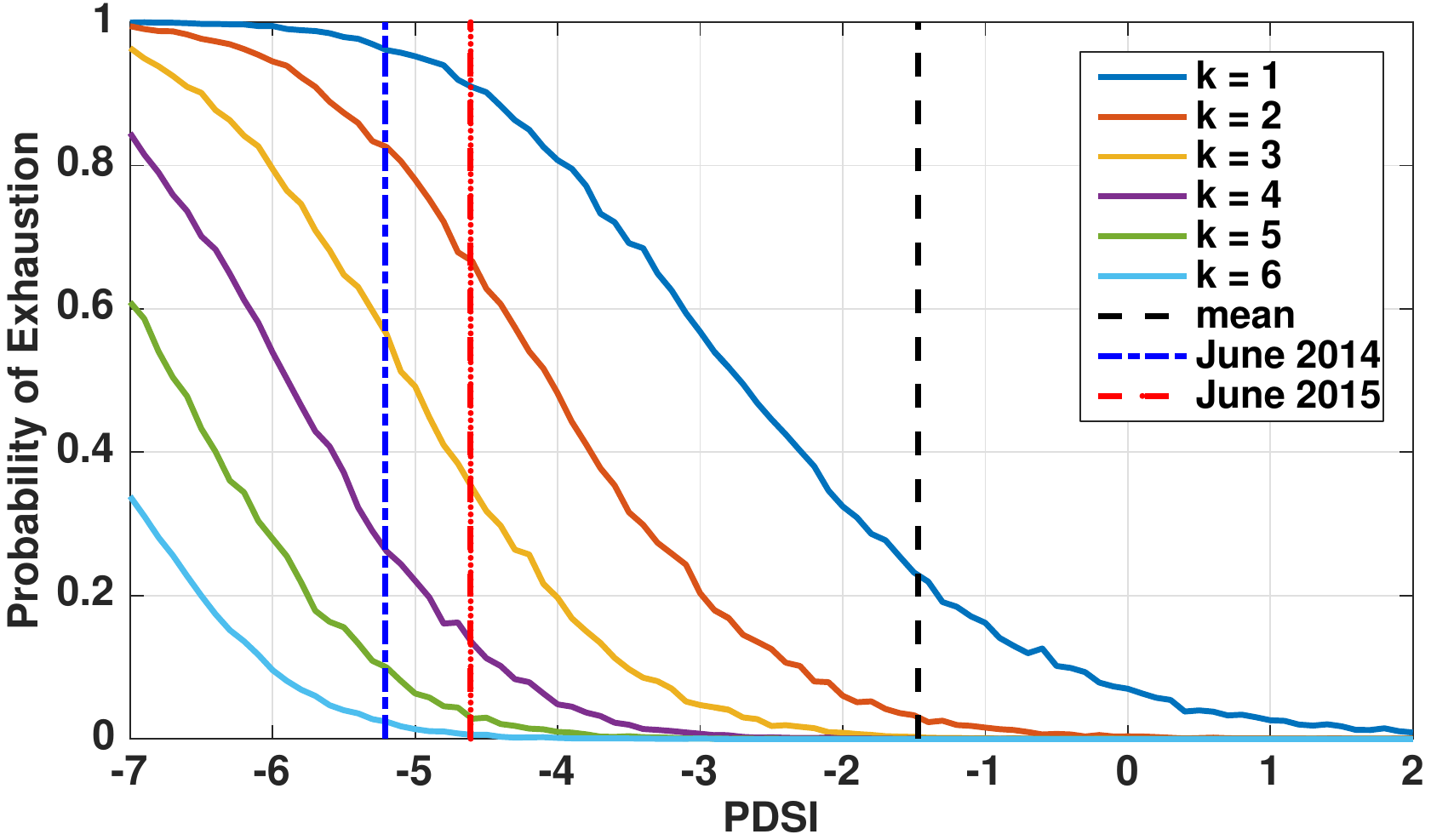}
\caption{Probability that at least $k$ reservoirs out of $31$ large reservoirs (with capacity $ \ge 10^{8}\mathrm{m}^3$) will have volume fall to zero, for a range of PDSI. 
Dashed black line: average June PDSI (June 2004--June 2015). Dashed blue line: June 2014 PDSI. Dashed red line: June 2015 PDSI.}
\label{fig:exhaustion1}
\end{figure}

Letting $\hat{\Sigma} = \hat{\Theta}^{-1}$, 
the composite variable $(y,x) \in \R^{56}$ 
is distributed as $(y,x) \sim \mathcal{N}(0,\hat{\Sigma})$.
(Preprocessing to remove climatology causes the mean to be zero.)
To determine the behavior of a collection of $K$ reservoirs 
${\bf r} = \{r_1,r_2,\dots,r_{K}\}$ as PDSI varies, 
we extract the $(K+1)\times (K+1)$ block of $\hat \Sigma$ 
corresponding to $y_{\bf r} \in \mathbb{R}^{K}$ and $x$, and recall that
\begin{equation}
  y_{\bf{r}} \given x 
  \sim
  \mathcal{N}(
    \hat{\Sigma}_{{y}_{\bf r},x} \hat{\Sigma}_{x}^{-1} x, \,
    \hat{\Sigma}_{y_{\bf r}} - \hat{\Sigma}_{y_{\bf r},x}\hat{\Sigma}_x^{-1}\hat{\Sigma}_{x,y_{\bf r}}
    )
    \quad,
\quad
\label{eq:exhaust-conditional}
\end{equation}
an instance of the standard expressions for the conditional mean
and variance of these jointly Gaussian variables. 
Let the August climatology, subtracted during preprocessing,
for reservoir volume $y_{r}$ ($r \in \mathbf{r}$) be $\mu_{y_{r}}$, and the June climatology of PDSI $x$ be $\mu_{x}$. Let the scaling used
to make the time series of $y_{r}$ and $x$ have unit variance
be $a_{y_r}$ and $a_x$. 
Then, for a June PDSI of $u$, 
the probability that at least 
$k$ of $K$ reservoirs have their volume drop below zero in August is:
\begin{equation}
  P(A_{K}(k) \given x = a_x(u - \mu_{x}))
  \label{eqn:jointexhaustion}
\end{equation}
where $A_{K}(k)$ is the event that $y_{r} \leq -\mu_{y_{r}}a_{y_{r}}$ for at least $k$ of the $K$ reservoirs. 
The probability in \eqref{eqn:jointexhaustion},
or that of any system-wide event, can be computed using 
Monte Carlo draws from the joint conditional distribution,
\eqref{eq:exhaust-conditional}. 
Here, we consider those reservoirs having capacity of at least $10^8\,\mathrm{m}^3$ 
($K = 31$)\footnote{Of the 55 reservoirs in our dataset, 22 have capacity below $10^8\,\mathrm{m}^3$.  Two of the 33 remaining (Terminus and Success) are flood-control reservoirs: they are unique in that their volume routinely falls below $10\%$ of capacity, independent of PDSI.  Thus, we focus on the remaining 31 large reservoirs in what follows.}. 
We vary PDSI and compute \eqref{eqn:jointexhaustion} for selected values of $k$.
Figure~\ref{fig:exhaustion1} indicates a high risk of simultaneous exhaustion for these major California reservoirs.

\begin{figure}[t!]
\centering
\bfseries{\textsf{~~ ~~Exhaustion Probabilities for Selected Reservoirs}}
\includegraphics[width=4.5in,height=2.2in]{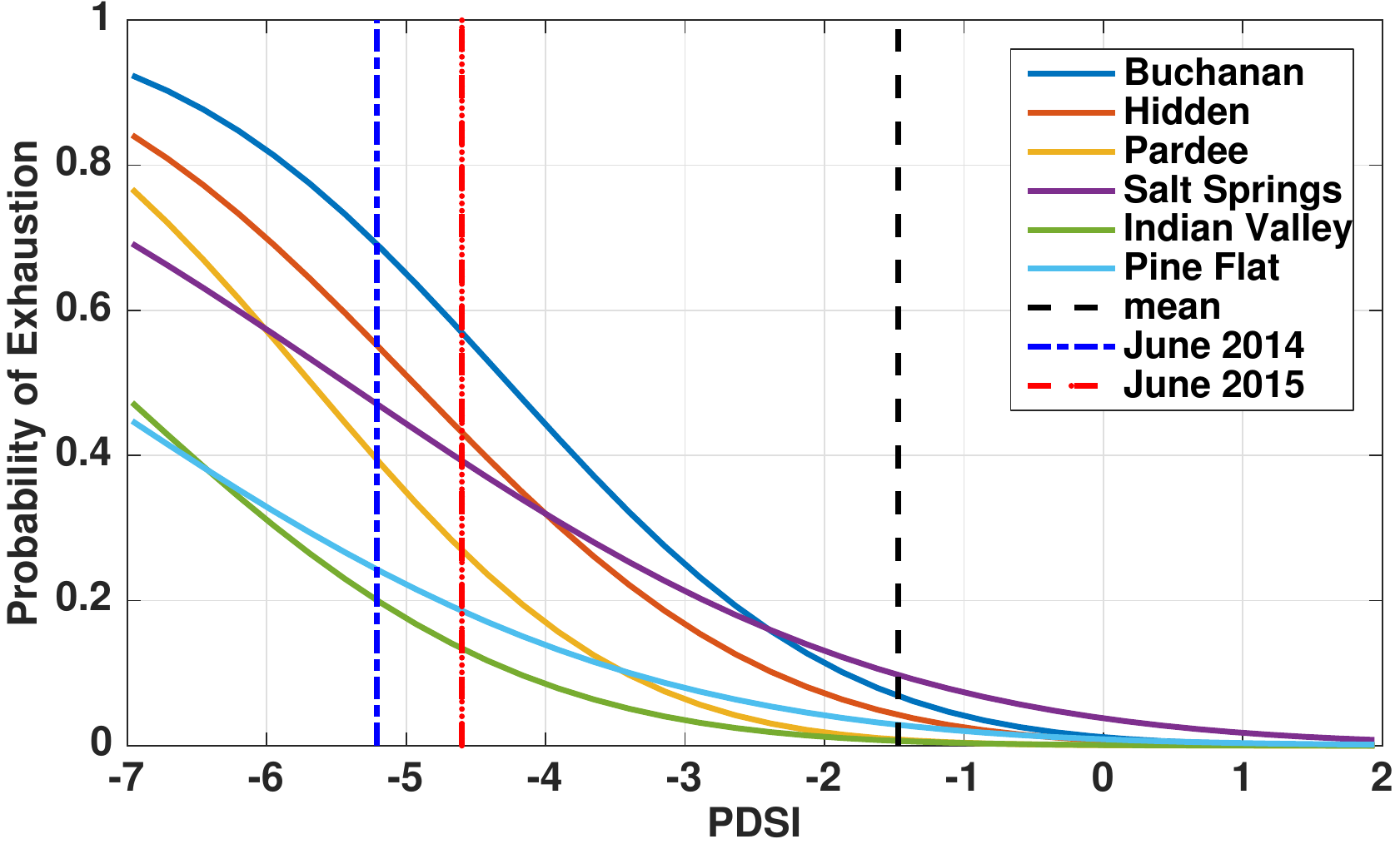}
\caption{Probability of exhaustion vs. PDSI for 6  large reservoirs 
(capacity $ \ge 10^{8}\mathrm{m}^3$)
at high risk of exhaustion. Dashed black line: average June PDSI (computed over June 2004-June 2015). Dashed blue line: June 2014 PDSI. Dashed red line: June 2015 PDSI.} 
\label{fig:exhaustion}
\end{figure}

\newpage
To identify
particular reservoirs that are at high risk of exhaustion, we compute the exhaustion probability 
of each reservoir conditioned on PDSI, namely,
\begin{equation}
  P(y_{r} < -\mu_{y_{r}}a_{y_{r}} \given x = a_x(u - \mu_{x}))
  \quad,
\end{equation}
by applying \eqref{eq:exhaust-conditional} with $K=1$.
For each reservoir, 
we sweep over a range of PDSI
to compute probabilities of exhaustion. 
Figure~\ref{fig:exhaustion} shows
those reservoirs (among 31 large reservoirs with capacity greater than $ 10^{8}\mathrm{m}^3$) that were highly sensitive to PDSI. Evidently, these reservoirs are at high risk of exhaustion,
and additionally, some have a greater sensitivity to small
PDSI changes than others.

\FloatBarrier

%
%

%% file: discuss.tex
%
%


\section{Discussion}
\label{sec:discuss}

We have used a representative subset of the full California reservoir network (55 of 1530 reservoirs) to model the relationships between reservoirs of different sizes and functions, and the correlation of system response to drought conditions. This unique approach allows us to produce an estimate of the global latent variables and discover that there are highly highly correlated with drought (PDSI) and hydropower generation. We deduce that PDSI, being computed from variables like precipitation and temperature that control mass balance, is a forcing function on system-wide reservoir levels, while correlation of water levels with aggregate hydropower generation is a system-wide response to high reservoir levels across the network.  While it is not surprising that PDSI is generally correlated with California reservoir levels, this reduced model allows for the first quantification of system-wide response to drought, a testing of the sensitivity to hypothetical conditions, and a highlighting of the most vulnerable elements of the reservoir network.

These results suggest that with sustained precipitation deficits and a PDSI approaching $-6$, the probability that four or more of California's major reservoirs (Figure~\ref{fig:exhaustion1}) run dry is greater than 50\%. This probability increase above 80\% as PDSI drops to $-7$. This does not imply that the reservoirs would in fact run dry.  Our approach assumes that the statistical distributions of reservoir response to PDSI changes are stationary, and would exhibit similar behavior under extreme PDSI values to that during the model training period.  Realistically, reservoir management behavior will change in response to the most extreme storage conditions. However, the model still demonstrates which reservoirs will be under the greatest threat if typical management behavior were to continue during extreme drought.  Additionally, the method used here can forecast other key events that precede reservoir exhaustion, such as when power generation is made impossible as water levels drop below turbine inlets.

The reservoir system's state is summarized by a complex, dynamic network of correlated time series that respond to a diverse set of global and local drivers, including both natural climate processes and human decision-making.  The experimental approach applied here to study reservoirs has the potential to be applicable across many complex data problems because it allows formation and testing of hypotheses about the suspected external drivers on a system, following a traditional experimental strategy.  The graphical modeling technique first reduced the number of connections between reservoirs, resulting in a robust model with high predictive power. Then a latent space summarizing all possible configurations of latent-variable times series was estimated by model optimization. Candidate external forcing time series were linked to this latent space to find matches. Once a best match is found (here according to correlation), then the latent variable could be included as a covariate in a new iteration of the graphical modeling procedure and an additional hypothesis for the existence of another latent variable could be tested.  This process can be repeated iteratively to identify and represent larger sets of drivers of global system variability. This approach also has value for directing or prioritizing observational efforts. 


%% file: Appendix.tex
\newpage
\section{Supporting Information}

\subsection{Linear Regression and Comparison with Graphical Modeling}
Linear regression is a commonly employed statistical technique for describing the relationship among a collection of variables in a hydrological setting. Applying this framework to our context, each reservoir volume anomaly
is modeled as a linear combination of other reservoirs. Thus, a given 
reservoir volume $y_r$ is modeled as a linear combination
\begin{equation}
y_r =  
  \begin{pmatrix} 
    y_{1} \, y_{2} \,\cdots\, y_{r-1} & y_{r+1} \,\cdots\, y_{55} 
  \end{pmatrix} 
  \beta_r + \epsilon,
\end{equation}
where $\beta_r \in \R^{54}$, and the 
scalar error $\epsilon$ is distributed as 
$\mathcal{N}(0,\sigma^2)$.
Given observations $\mathcal{D}_n = \{y^{(i)}\}_{i = 1}^n \subset \R^{55}$, 
the maximum likelihood estimator 
for the parameter $\beta_r$ solves the following optimization program:
\begin{equation}
 \hat{\beta}_r 
 = 
 \argmin_{\beta_r \in \R^{54}} 
   \| Y_r - \tilde{Y}_r \beta_r \|_{2}^2,
\label{eqn:LinearRegression}
\end{equation}
where $\| \cdot \|_2$ is the euclidean norm, $Y_r = \begin{pmatrix} y_r^{(1)} \,  y_r^{(2)} \dots y_r^{(n)} \end{pmatrix}^T$, and \\ $\tilde{Y}_r =  \begin{pmatrix} y_1^{(1)} & y_2^{(1)} & \cdots y_{r-1}^{(1)} & y_{r+1}^{(1)} & \cdots y_{55}^{(1)} \\ \vdots & \vdots&  \vdots &  \vdots & \vdots \\ y_1^{(n)} & y_2^{(n)} & \cdots y_{r-1}^{(n)} & y_{r+1}^{(n)} & \cdots y_{55}^{(n)} \end{pmatrix}$.
Given sufficient observations,
\eqref{eqn:LinearRegression} has a closed-form solution:
\begin{equation}
  \hat{\beta}_r = (\tilde{Y}_r^T \tilde{Y}_r)^{-1} \tilde{Y}_r^T Y_r
  \quad.
\label{eqn:Lin}
\end{equation}
This procedure can be repeated for every $r \in \{1,2, \dots 55\}$ to obtain a complete model. \\

We now discuss advantages of using graphical modeling over classical linear regression for modeling this reservoir system. We use $\mathcal{D}_n$ as input to \eqref{eqn:Lin} to learn a full linear regression model. We compare this model with the graphical model obtained in Section~\ref{sec:results}.  We evaluate the performance of each model via the \emph{coefficient of determination} ($R^2$) which measures the fraction of variance of each reservoir explained by the model, i.e.:
\begin{equation}
    R^2 = 1 - \frac{SS_{res}}{SS_{tot}}
    \label{eqn:R^2}
\end{equation}
For each reservoir $y_r$, we compute the coefficient of determination specified by our model. Let $\bar{y}_r = \frac{1}{n_\train} \sum_{i=1}^{n_\train} y_{r}^{(i)}$ denote the average volume of reservoir $y_r$ over the training set. Furthermore, let $f_{r}^{(i)}$ denote a model's prediction of reservoir volume $y_r$ at a specified time instance. Then the quantities $SS_{tot}$ and $SS_{res}$ in \eqref{eqn:R^2} are given by $SS_{tot} = \sum_{i = 1}^{n_\test} (y_r^{(i)}-\bar{y}_r)^2$ and $SS_{res} = \sum_{i = 1}^{n_\test}(f_{r}^{(i)}-\bar{y}_r)^2$ respectively. We evaluate the quantity $f_{r}^{(i)}$ by computing the expected volume of reservoir $y_r$ conditioned on the volumes of the remaining reservoirs. For the linear regression model with the estimate $\hat{\beta}_r$, $f_r^{(i)}$ is equal to
\begin{eqnarray*}
f_{r}^{(i)} = \begin{pmatrix} y_1^{(i)}, y_2^{(i)}, \dots y_{r-1}^{(i)}, y_{r+1}^{(i)}, \dots y_{55}^{(i)} \end{pmatrix}\hat{\beta}_r. 
\end{eqnarray*}
Let $\hat{\Theta}$ denote the estimated precision matrix of the graphical model and $\hat{\Sigma} = \hat{\Theta}^{-1}$ the corresponding covariance matrix. Further, let $\hat{\Sigma}_{r}$ denote the extracted $54 \times 54$ covariance matrix among reservoirs $\begin{pmatrix} y_1,y_2,\dots y_{r-1}, y_{r+1}, \dots y_{55} \end{pmatrix}$, and \\
$\sigma_{r} = \begin{pmatrix} \Sigma_{y_1,y_r},\Sigma_{y_2,y_r},\dots, \Sigma_{y_{r-1},y_r}, \Sigma_{y_{r+1},y_r}, \dots, \Sigma_{y_{55},y_r} \end{pmatrix}$ denote a row vector of cross covariances between reservoir $y_r$ and the other $54$ reservoirs. Then the quantity $f_{r}^{(i)}$ corresponding in the graphical model is
\begin{eqnarray*}
f_{r}^{(i)} = \sigma_r\hat{\Sigma}_{r}^{-1}\begin{pmatrix} y_{1}^{(i)}, y_{2}^{(i)},\dots y_{r-1}^{(i)}, y_{r+1}^{(i)}, \dots y_{55}^{(i)} \end{pmatrix}^T.
\end{eqnarray*}

\begin{figure}[!http]
\centering
\includegraphics[width=4in]{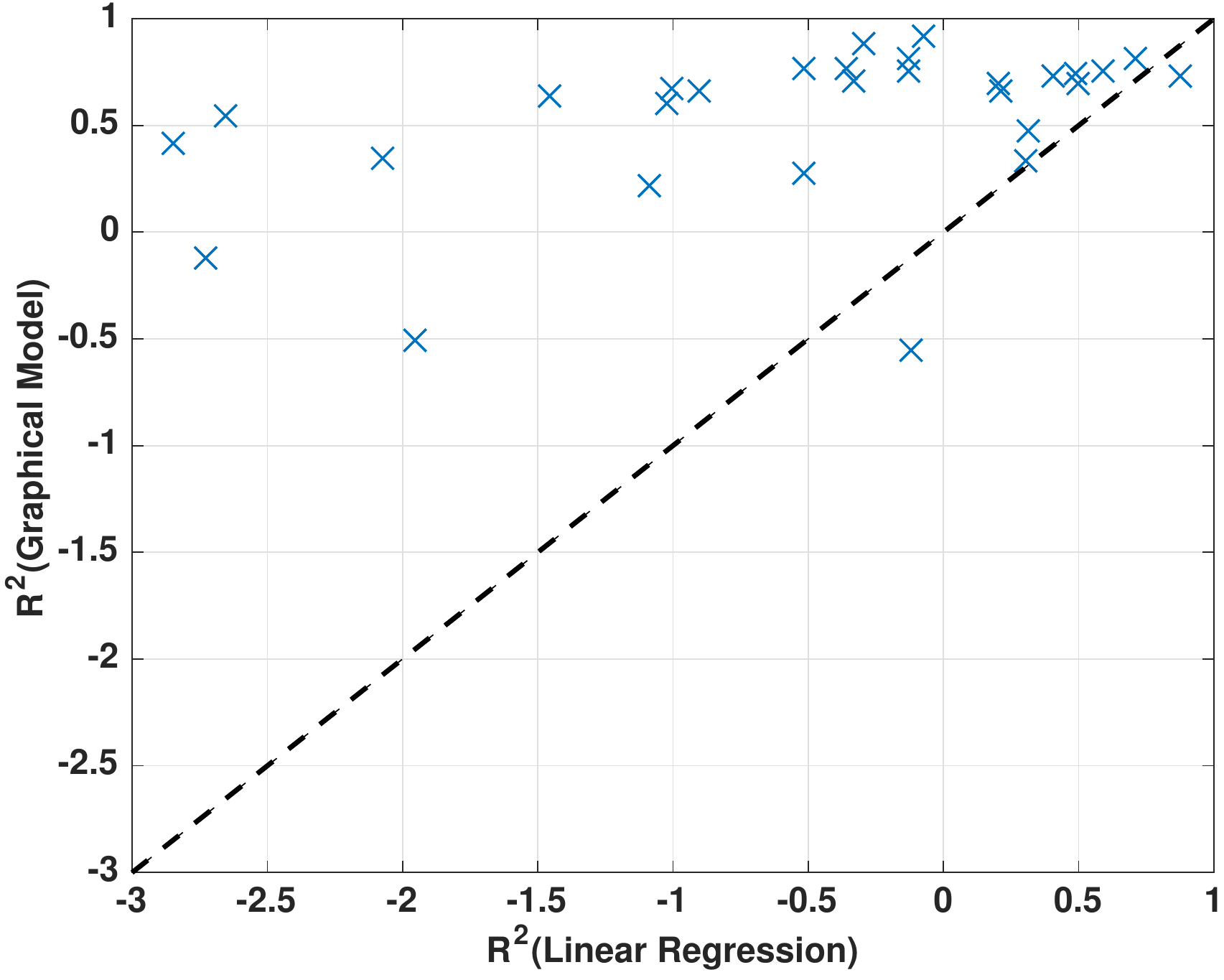}
\caption{$R^2$ comparison of a model learned via linear regression and a model learned via graphical modeling. Each cross denotes a reservoir. Crosses above the dotted line suggests that graphical modeling is better suited for modeling the variation of the given reservoir.}
\label{fig:R^2comparison}
\end{figure}

Figure~\ref{fig:R^2comparison} compares the $R^2$ values corresponding to the model obtained via linear regression and the model obtained via graphical modeling. Here, only a subset of reservoirs are plotted as linear regression results in very negative $R^2$ values for the excluded subset. We observe that for a large portion of the reservoirs, linear regression yields negative $R^2$ values. This exemplifies the danger of over-fitting. On the other the hand, excluding one reservoir, graphical modeling produces strictly larger $R^2$ values, indicating that graphical modeling is better suited to modeling this reservoir system.
\FloatBarrier
\FloatBarrier

\subsection{Conditional Latent-Variable Graphical Modeling}
In this section, we present an estimator for learning a conditional latent-variable graphical modeling. As described in Section~\ref{sec:method}, the conditional latent-variable graphical modeling allows for incorporating covariate vectors $x \in \R^q$. Let $\Sigma \in \Sp^{55+q}$ be the joint covariance matrix over reservoir volumes and covariates $(y,x) \in \R^{55+q}$. A conditional latent-variable graphical model specifies the distribution of $y | x$ as a latent-variable graphical model. Following the description of the latent-variable modeling framework in Section~\ref{sec:method}, this implies that the conditional precision matrix of $y$ given $x$ must be decomposable as the difference of a sparse and a low-rank matrix. This algebraic property can be naturally stated using the joint precision matrix $\Theta = \Sigma^{-1}$ with conditions on the submatrix $\Theta_y$ of $\Theta = \begin{pmatrix} \Theta_y & \Theta_{yx} \\ \Theta_{yx}' & \Theta_{x} \end{pmatrix}$. 
The conditional precision matrix of $y$ given $x$ is equal to the submatrix $\Theta_y$; thus $\Theta_y$ is the difference of a sparse and low-rank matrix. Based on this algebraic structure desired on $\Theta$, the following is a natural convex relaxation for fitting the augmented data $\data_n^{+} = \{ (y^{(i)},\, x^{(i)})\} \subset \R^{55+q}$ to this conditional latent-variable graphical model:

\begin{eqnarray}
  (\hat{\Theta}, \hat{S}_y, \hat{L}_y)  = 
  \argmin_{\substack{\Theta \in \Sp^{55+q} \\ S,L \in \Sp^{55}}} 
  &\eqnarrpad&
  -\ell(\Theta; \data_n^{+}) 
  + 
  \lambda ( \gamma \|S_y\|_{1} + \tr(L_y) ) \nonumber 
  \\ 
  \suchthat 
  &\eqnarrpad&
  \Theta \succ 0, ~ \Theta_y = S_y - L_y, ~ L_y \succeq 0.
\label{eqn:composite}
\end{eqnarray}
The term $\ell(\Theta; \data_n^{+})$ 
is the (scaled) Gaussian log-likelihood function over 
the variables $(y,x)$, given by
\begin{equation}
    \ell(\Theta; \data_n^{+}) 
    = 
    \log\det(\Theta) - 
    \tr \left[\Theta \cdot \tfrac{1}{n}\sum_{i=1}^n \begin{pmatrix} y^{(i)} \\ x^{(i)} \end{pmatrix} \begin{pmatrix}y^{(i)} \\ x^{(i)} \end{pmatrix}'\right]
\quad,
\end{equation}
where we have again dropped constant terms and scaled by $-2$.

%% file: main.bbl
\begin{thebibliography}{99}
\expandafter\ifx\csname natexlab\endcsname\relax\def\natexlab#1{#1}\fi

\bibitem{aghakouchak2014global}
\textsc{AghaKouchak, A, Cheng, L., Mazdiyasni, O. $\&$ Farahmand, A.}
\newblock{Global warming and changes in risk and concurrent climate extremes: Insights from the 2014 California drought}.
\newblock \textit{Geophysical Research Letters}. 41:8847--8852, 2014.

\bibitem{barnett:dry-2008}
\textsc{Barnett, T. $\&$ Pierce, D.}
\newblock{When will {L}ake {M}ead go dry?}.
\newblock\textit{Water Resources Research}. 44, 2008.

\bibitem{barnett2009sustainable}
\textsc{Barnett, T. $\&$ Pierce, D.}
\newblock{Sustainable water deliveries from the Colorado River in a changing climate}.
\newblock\textit{Proceedings of the National Academy of Sciences}. 106:7334--7338, 2009.

\bibitem{Chand2012}
\textsc{Chandrasekaran, V., Parrilo, P. A. $\&$ Willsky, A. S.}
\newblock {{Latent Variable Graphical Model Selection via Convex Optimization}}.
\newblock \textit{Annals of Statistics}, 40:1935--1967, 2012.

\bibitem{diffenbaugh2015anthropogenic}
\textsc{Diffenbaugh, N., Swain, D., $\&$ Touma, D.}
\newblock{Anthropogenic warming has increased drought risk in California}.
\newblock\textit{Proceedings of the National Academy of Sciences}. 112:3931--3936, 2015.

\bibitem{Fazel}
\textsc{Fazel, M.}
\newblock \textit{{Matrix rank minimization with applications}}.
\newblock {PhD thesis, Dept. Elec. Engr., Stanford
University}, 2002.

\bibitem{famiglietti2014global}
\textsc{Famiglietti, J.}
\newblock{The global groundwater crisis}.
\newblock\textit{Nature Climate Change}. 4:945--948, 2014.

\bibitem{Friedman}
\textsc{Friedman, J., Hastie, T. $\&$ Tibshirani, R.}
\newblock {{Sparse inverse covariance estimation with the graphical lasso}}.
\newblock \textit{Biostatistics}, 9:432-441, 2008.

\bibitem{graf1999dam}
\textsc{Graf, W.}
\newblock{Dam nation: A geographic census of American dams and their large-scale hydrologic impacts}.
\newblock\textit{Water Resources Research}. 35:1305--1311, 1999.

\bibitem{griffin2014unusual}
\textsc{Griffin, D. $\&$ Anchukaitis, K.}
\newblock{How unusual is the 2012--2014 California drought?}
\newblock\textit{Geophysical Research Letters}. 41:9017--9023, 2014.

\bibitem{hoell2015does}
\textsc{Hoell, A., et al.}
\newblock{Does El Ni{\~n}o Intensity Matter for California Precipitation?}
\newblock\textit{Geophysical Research Letters}. 43:819--825, 2015.

\bibitem{howitt2014economic}
\textsc{Howitt, R., Medell{\'\i}n-Azuara, J., MacEwan, D., Lund, J. $\&$ Sumner, D.}
\newblock{Economic analysis of the 2014 drought for California agriculture}.
\newblock{Center for Watershed Sciences, University of California, Davis}, 2014.


\bibitem{shukla2015temperature}
\textsc{Shukla, S., Safeeq, M., AghaKouchak, A., Guan, K. $\&$ Funck, C.}
\newblock{Temperature impacts on the water year 2014 drought in California}.
\newblock\textit{Geophysical Research Letters}.42:4384--4393, 2015.

\bibitem{solander2016simulating}
\textsc{Solander, K., Reager, J., Thomas, B., David, C. $\&$ Famiglietti, J.}
\newblock{Simulating human water regulation: {T}he development of an optimal complexity, climate-adaptive reservoir management model for an LSM}.
\newblock\textit{Journal of Hydrometeorology}. 17:725--744, 2016.

\bibitem{Toh}
\textsc{Toh, K. C, Todd, M. J. $\&$ Tutuncu, R. H.}
\newblock {SDPT3 - A
MATLAB software package for semidefinite-quadratic-linear
programming}. 
\newblock Available from
\url{http://www.math.nus.edu.sg/~mattohkc/sdpt3.html}.
Accessed April 2016.

\bibitem{Wainwright}
\textsc{Wainwright, M. J.}
\newblock{Structured regularizers for high-dimensional problems: Statistical and computational issues}.
\newblock \textit{Annual Review of Statistics and its Applications}. 1:233--253, 2014.

\bibitem{williams2015contribution}
\textsc{Williams, A., et al.}
\newblock{Contribution of anthropogenic warming to California drought during 2012--2014}.
\newblock{\textit{Geophysical Research Letters}}, 42:6819--6828, 2015.

\bibitem{Yuan}
\textsc{Yuan, M. $\&$ Lin, Y.}
\newblock{Model selection and estimation in the Gaussian graphical model}.
\newblock{Biometrika}, 94:19--35, 2007.

\end{thebibliography}
